\documentclass[preprint,showpacs,preprintnumbers,amsmath,amssymb]{revtex4}

\pdfoutput=1

\usepackage{graphicx}
\usepackage{bm}

\begin{document}

\title{Design of Optical Tunable CNOT (XOR) and XNOR Logic Gates Based on $2$D-Photonic
Crystal Cavity Using Electro-Optic Effect}

\author{Karim Abbasian}
\affiliation{The School of Engineering Emerging Technologies,
University of Tabriz, Tabriz, 51664, Iran}%
\author{Rasool Sadeghi}
\affiliation{Department of Electrical Engineering, Ahar branch, Islamic Azad University, Ahar, Iran}%
\author{Parvin Sadeghi}
\affiliation{ Marand Faculty of Engineering, University of Tabriz, Tabriz, Iran.}%
\email{psadeghi@tabrizu.ac.ir}

\date{\today}

\begin{abstract}
We have proposed optical tunable CNOT (XOR) and XNOR logic gates
using two-dimensional photonic crystal ($2D PhC$) cavities. Where,
air rods with square lattice array have been embedded in Ag-Polymer
substrate with refractive index of $1.59$. In this work, we have
enhanced speed of logic gates by applying two input signals with a
phase difference at the same wavelength for $2D PhC$ cavities.
Where, we have adjusted the phases of input and control signals
equal with $\pi/3$ and zero, respectively. The response time of the
structure and quality factor of the cavities are in the range of
femtosecond and $2000$, respectively. Then, we have used
electro-optic property of the substrate material to change the
cavities resonance wavelengths. By this means, we could design the
logic gates and demonstrate a tunable range of $23nm$ for their
operation wavelength. The quality factor and the response times of
cavities remain constant in the tunable range of wavelength,
approximately. The evaluated least ON to OFF logic-level contrast
ratios for the XOR and XNOR logic gates are $25.45 dB$ and $22.61
dB$, respectively. The bit rates of the proposed logic gates can
reach up to higher than $0.166$ $P(Peta)bps$ values. According to
the high rapid response time with acceptable quality factor of the
$PhC$ cavities, the proposed optical logic gates can be considered
as appropriate candidates to be building blocks for applications
such as optical integrated circuits and optical processors with an
ultrahigh speed of data flow.

\end{abstract}

\maketitle

\section{Introduction}
According to the reported works to date, in information processors
and communication systems, employing ultra-fast and cheap components
provide them with competitive ability in market challenges. As a
candidate to achieve this goal, $2D PhC$ based all-optical devices
(all optical logic gates) has led to significant increment in
operating speed and consequently in the performance bandwidth of the
systems. However, according to the recent reports, the response time
of these logic gates (more of them) has been enhanced to picoseconds
(ps) range \cite{Fushimi2014, Christina2012, Noshad2007, Liu2008}.
Fushimi \textit{et al.} designed scalable all-optical logic gates
with the same wavelength of the input and output signals. They
allowed a wavelength fluctuation, with $0.3$ fraction of the cavity
resonant wavelength widths. Then, they investigated cavities
coupling to the waveguide and the operation degradation of logic
gates. Also, Fushimi investigated the scalability of the designed
logic gates considering their input power stability and tolerable
fabrication errors \cite{Fushimi2014}. Christina \textit{et al.}
proposed all optical AND, NAND, XNOR and NOR logic gates based on
$2D PhC$s using self-collimation effect in a hexagonal structure by
creating line defect in the structure \cite{Christina2012}. Noshad
\textit{et al.} proposed AND, NOR and NOT all optical logic gates
based on $2D PhC$s. They changed operational wavelength of the
structure by embedding Kerr nonlinear rods in the structure
\cite{Noshad2007}. Liu \textit{et al.} proposed an all optical half
adder based on cross structure $2DPhCs$. They first designed AND and
XOR logic gates and then used them to make a half adder. They first
designed an AND gate in a nonlinear structure and an XOR gate in
another nonlinear structure and then combined them to make a half
adder. Furthermore, they calculated an optimal operation speed
without considering response time of the Kerr nonlinear material in
the $ps$ temporal range \cite{Liu2008}.

Using semiconductor micro-ring resonators is another alternative
approach for all optical logic functions realizing
\cite{Isfahani2009,Lin2013}. Bai \textit{et al.} proposed all
optical NOT and NOR logic gates by using a ring resonator. In this
work they controlled the input signals by applying an optical probe
wave to the ring. Where, wavelength of the probe and input signals
were the same. Their logic gates were based on $2D PhC$s
\cite{Bai2009}. To the best of our knowledge, in the most of
reported works, silicon rods have been used in the air or silica
substrates but in among them some researchers have used ring
resonators to perform logic gate functions. As instance, Andalib
\textit{et al.} could propose a controllable compact logic gate by
using nonlinear ring resonators based on $2D PhC$s
\cite{Andalib2008}. They could also propose all optical AND and NOR
logic gates by their proposed nonlinear ring resonator
\cite{Andalib20081, Andalib2009}. In spite of their claim, the
proposed gates were big in size and consequently had low operational
speed. Also, other alternative methods such as all optical logic
gate designing by using ultra small $PhC$ heterojunction diodes have
been reported \cite{Wang2013}. They analyzed all-passive on-chip
optical AND and NAND logic gates made from a directional emitting
cavity connecting two ultra small $PhC$ heterojunction diodes.
Where, the analyzed logic gates were phase insensitive. However,
proposed logic gates by Fu \textit{et al.} were depend on phase of
the input signals that created along the wave guides at the $2D PhC$
structure \cite{Fu2012}.

Liu \textit{et al.} proposed ultra-fast all optical AND, NAND, OR
and NOR logic gates by using cavities based on $2D PhC$s, which
operates with low input power and very low response times in around
of femtosecond ($fs$) \cite{liu2011}. In this work, they utilized
air rods in the Ag-Polymer substrate. By using the nonlinear
property of Ag-Polymer they shifted the cavities resonance
wavelength and thereby could propose all optical logic gates.
According to other studies, the structures made of Polyester based
on $2D PhC$s have ultra-fast response times ($fs$) \cite{liu2009,
liu2005, Hu2005}.

On the other hand, tunable $PhC$ components are attractive subjects
in the field of the modern optical communication systems extension.
Tunable electro optical devices based on $2D$ and also $3D PhC$
structures have been proposed \cite{Liu2004, Liu20051, Liu20041,
Liu20052, Liu2006}. The experimental proof of an electro-optical
$PhC$ was investigated and found to be in agreement with numerical
results \cite{Amet2010}. Although photonic crystals can be employed
to achieve low group velocities at their band edges, this is limited
to a very narrow range of wave vectors in one particular direction.
Recently, two-dimensional arrays of coupled photonic crystal
resonators have been a study focus which exhibit reduced group
velocities over the entire range of wave vectors \cite{Altug2005,
Moreolo2008}. It is well known that the electro-optic effect has an
ultra low response time of the order of nanosecond. This property
makes it highly desirable and opens up the possibilities for ultra
fast tunable $PhC$ devices with low power. Yang \textit{et al.} have
investigated a novel-designed two-dimensional coupled photonic
crystal resonator array ($2D CPCRA$) realized in $2D PhC$ slabs
filled with nonlinear polymer, which can dynamically tune the slow
light properties in a wide frequency range and realize optical
devices that could store and release optical pulses to implement
values optical digital processing in future high speed optical
networks \cite{Yang2011}.

In the present work, we used $2D PhC$s based cavities composed of
air rods created in the Ag-polymer. It should be noted that we did
not consider nonlinear properties of substrate (Ag-Polymer). The
created phase shifts by the cavities was employed to propose all
optical CNOT (XOR) and XNOR logic gates. This paper is composed of
three sections. In the first section, we demonstrate and describe
cavity, created in the 2DPhC, and the cavities resonance wavelengths
with investigation of electro-optical effects of that on the
cavities. In the second section, simulation results and their
discussions have been presented and the   final section is devoted
to the conclusion and the obtained research results.

\section{Design and realization of structure}
The employed cavity is created in a $2D PhC$ including air rods with
$0.3a$ radii perforated in the Ag-Polymer substrate with refractive
index of $1.59$ and lattice constant of $a = 550nm$. By removing of
some central air rods a L-type cavity is created, where removing of
seven or eight rods from a single row of the $PhC$ results in $L7$
or $L8$ cavities, respectively. By decreasing radii of a line of
nearest-neighbor rods of the cavity and also shifting them away, one
can enhance the resonance wavelength ($\lambda_r$) and quality
factor ($Q$) of the cavities. Here, we have decreased radii of a
line of nearest-neighbor rods of the cavity to the value $0.25a$ and
also shifted them and the second neighbor rods away in about $0.19a$
and $0.1a$, respectively. In table $1$ we show the values of $Q$ and
$\lambda_r$ for the cavities. Also, the schematic structure of
cavity is depicted in Figure $1$.

\begin{table}
  \centering
  \caption{Resonance wavelength and quality factor for two cavity types without any voltage bias.}
  \begin{tabular}{ccccc} \\ \hline
   Cavity type & Resonate wavelength ($\lambda_r$) & Quality factor ($Q$) \\ \hline
   $L7$ & $1561.1nm$ & $1472.40$ \\
    $L8$ & $1562.7nm$ & $1816.86$ \\ \hline
  \end{tabular}

  \centering

\end{table}

\begin{figure}[htb]
\begin{center}
\centerline{\includegraphics[height=6cm]{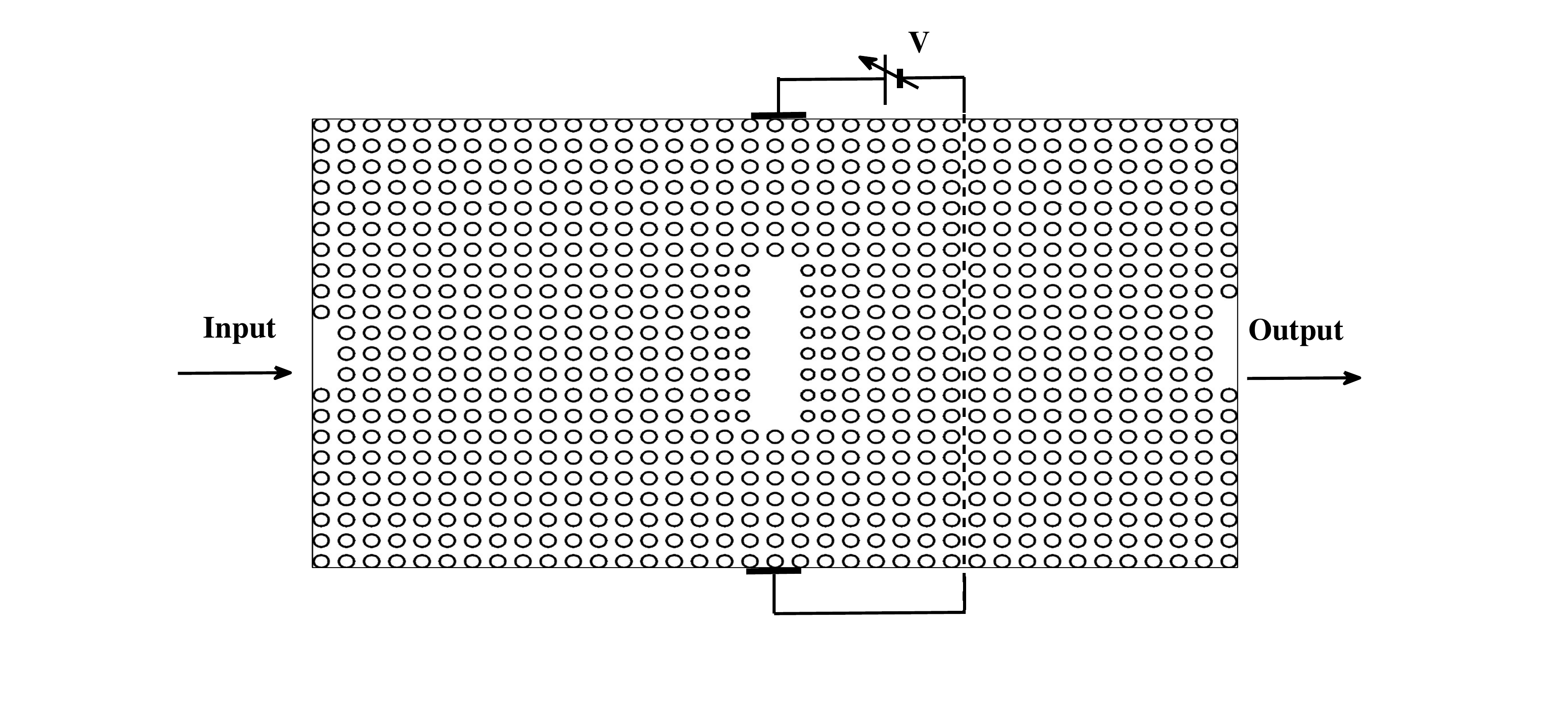}} \caption{(Color online)
Schematically view of the L-type cavity structure in two-dimensional
photonic crystals.} \label{Fig. 1.}
\end{center}
\end{figure}

\begin{figure*}[htb]
\centerline{
\includegraphics[height=6cm]{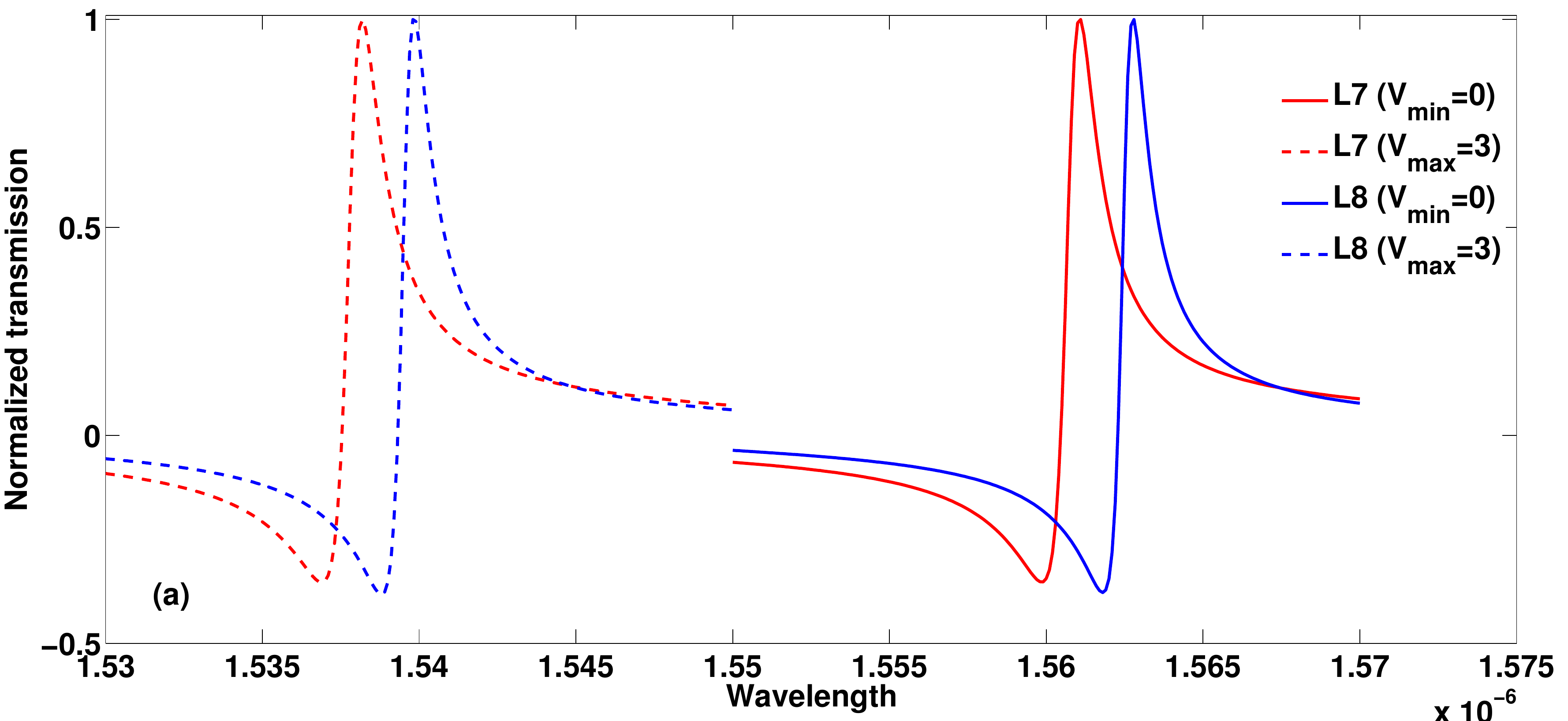}}
\centerline{
\includegraphics[height=3cm]{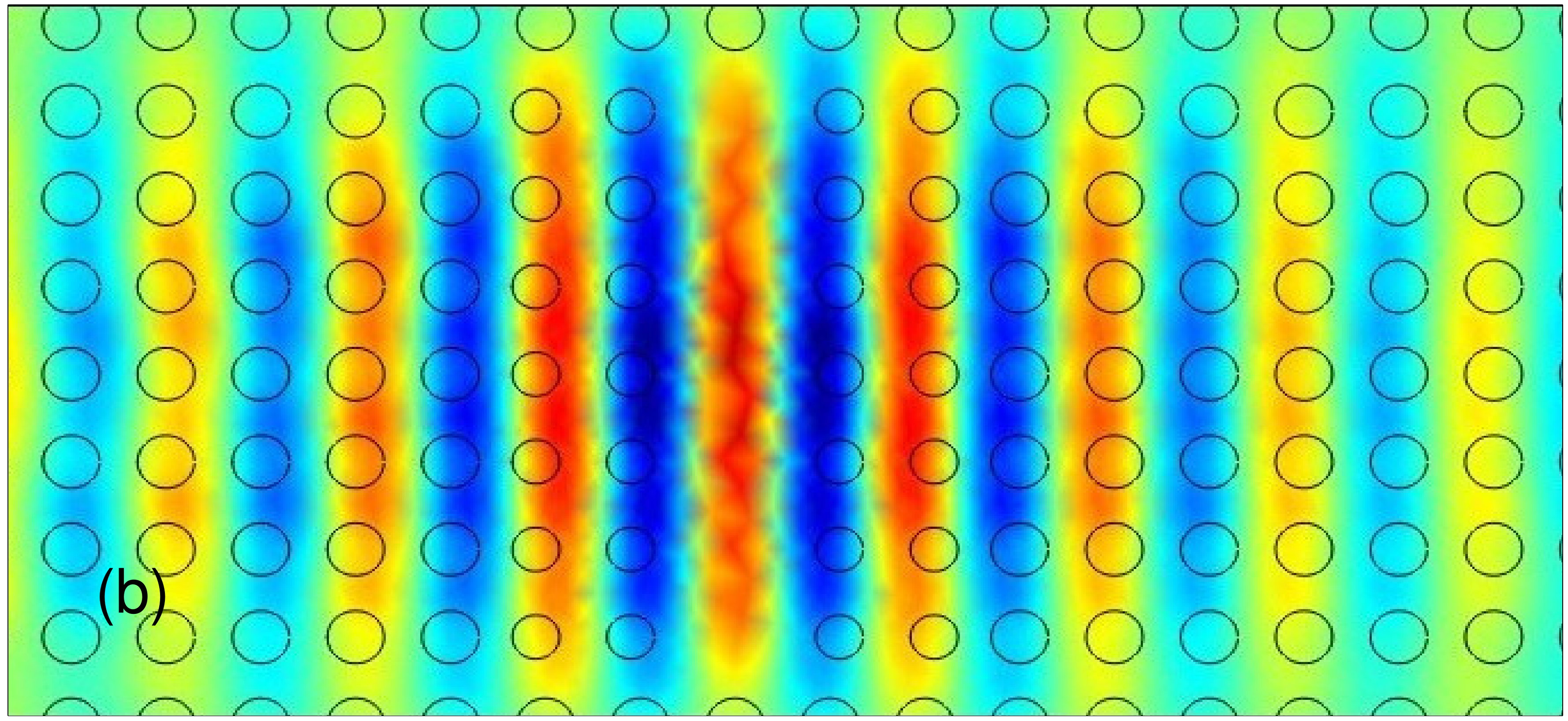}
\includegraphics[height=3cm]{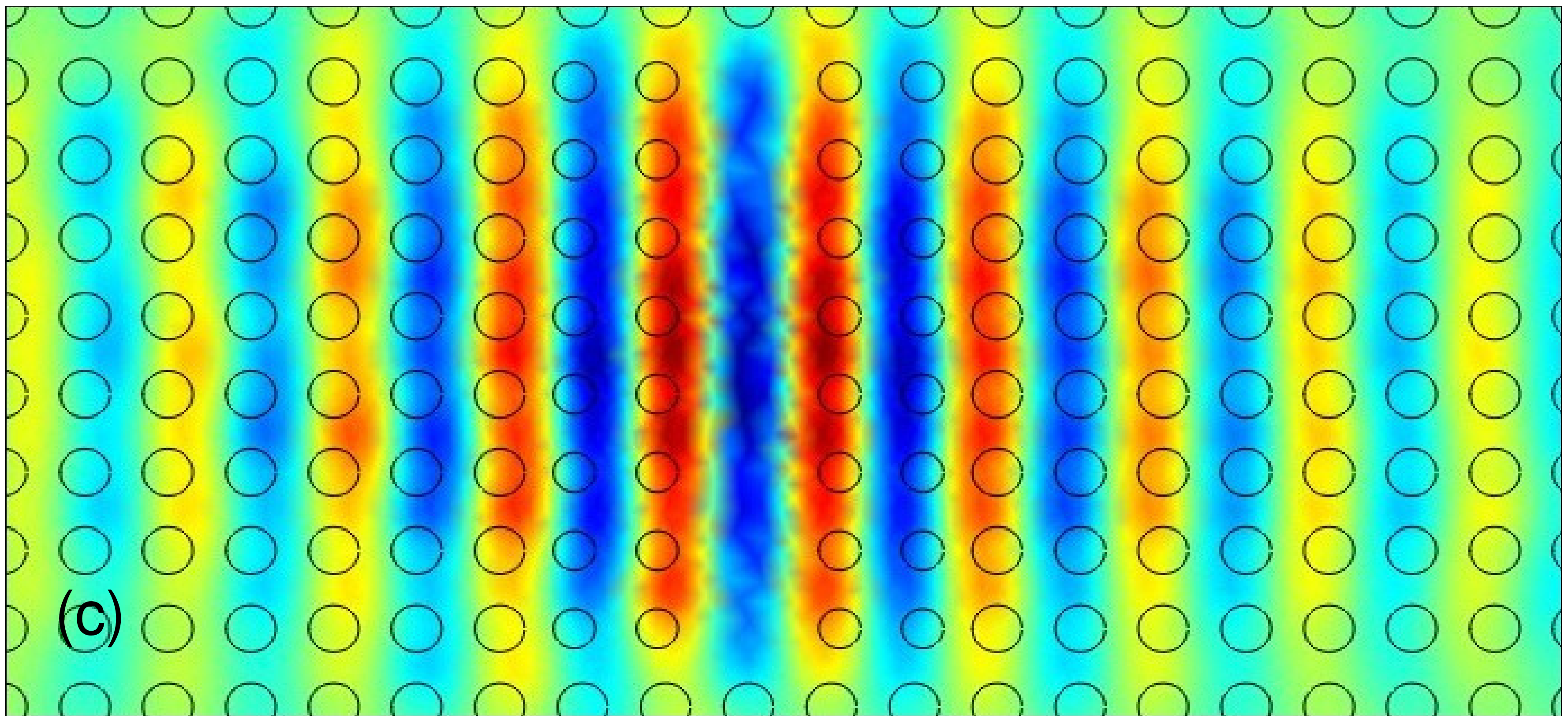}}
\caption{(Color online) a) Normalized transmission spectrum profile
for the $L7$ and $L8$ cavities. b)Field distribution of the $L7$
cavity at wavelength of $1561.8 nm$. c) Field distribution of the
$L8$ cavity at a wavelength in the tunable rang from $1561.8 nm$ to
$1538.8 nm$.} \label{Fig. 2.}
\end{figure*}

\begin{figure}[htb]
\begin{center}
\includegraphics[height=6cm]{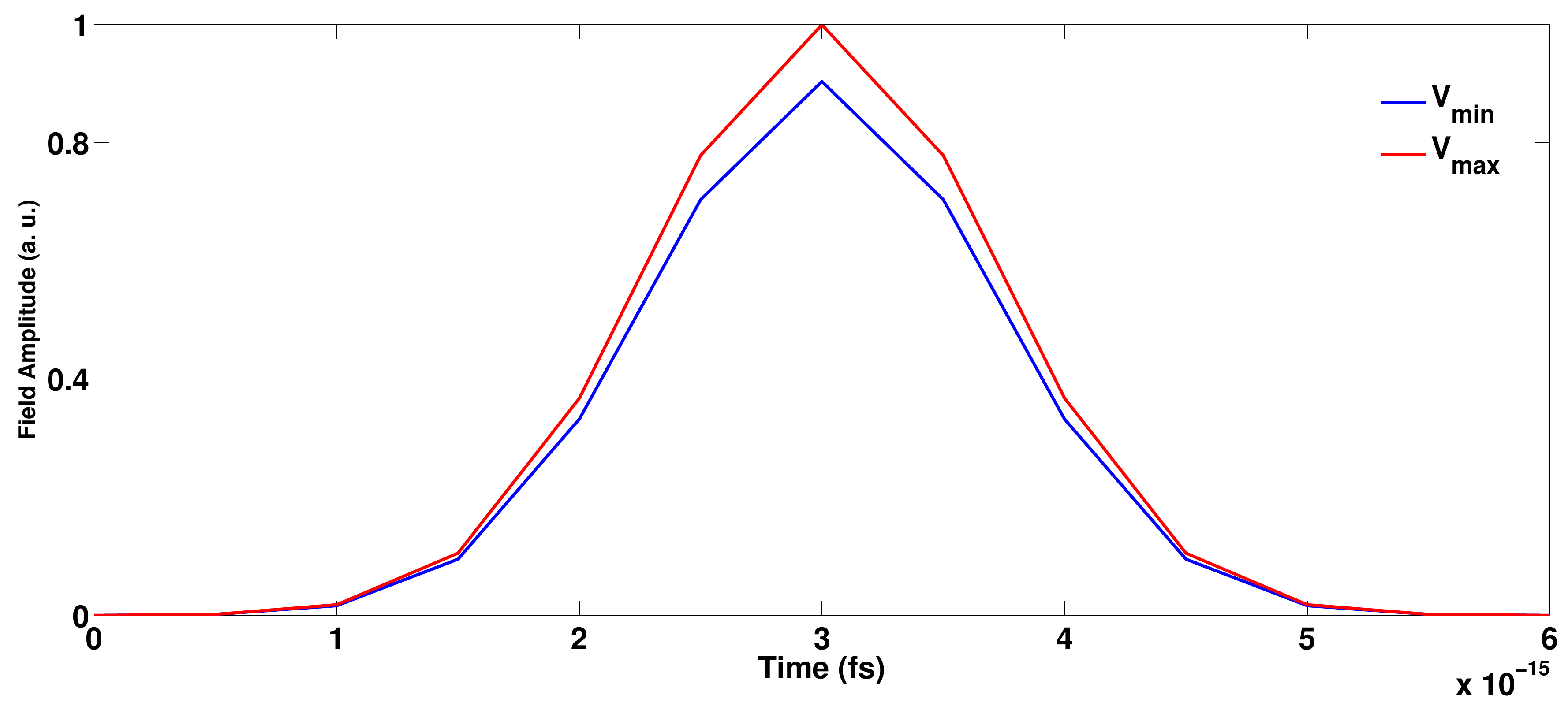}
\caption{(Color online) Temporal view of the output pulse. }
\label{fig4}
\end{center}
\end{figure}

\begin{figure*}[htb]
\centerline{
\includegraphics[height=3cm]{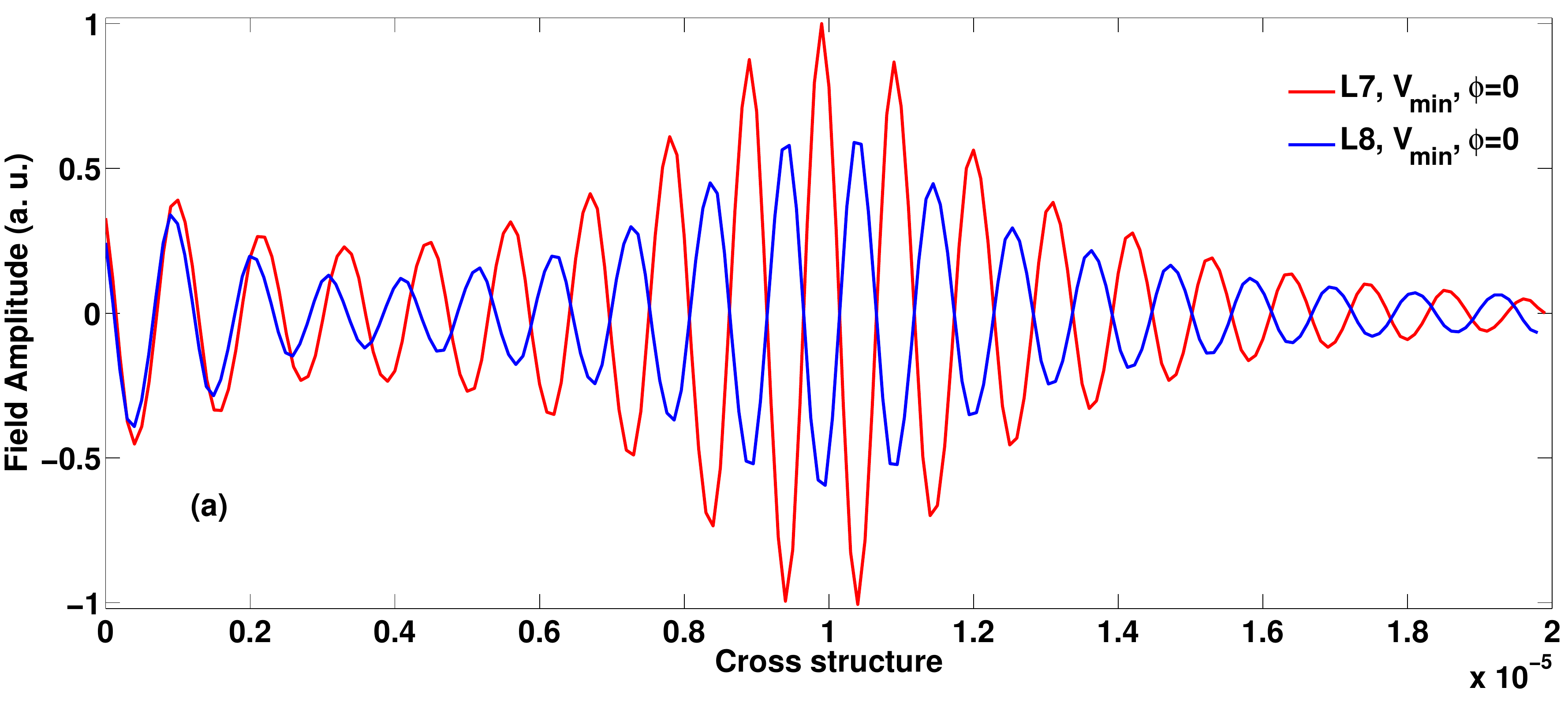}
\includegraphics[height=3cm]{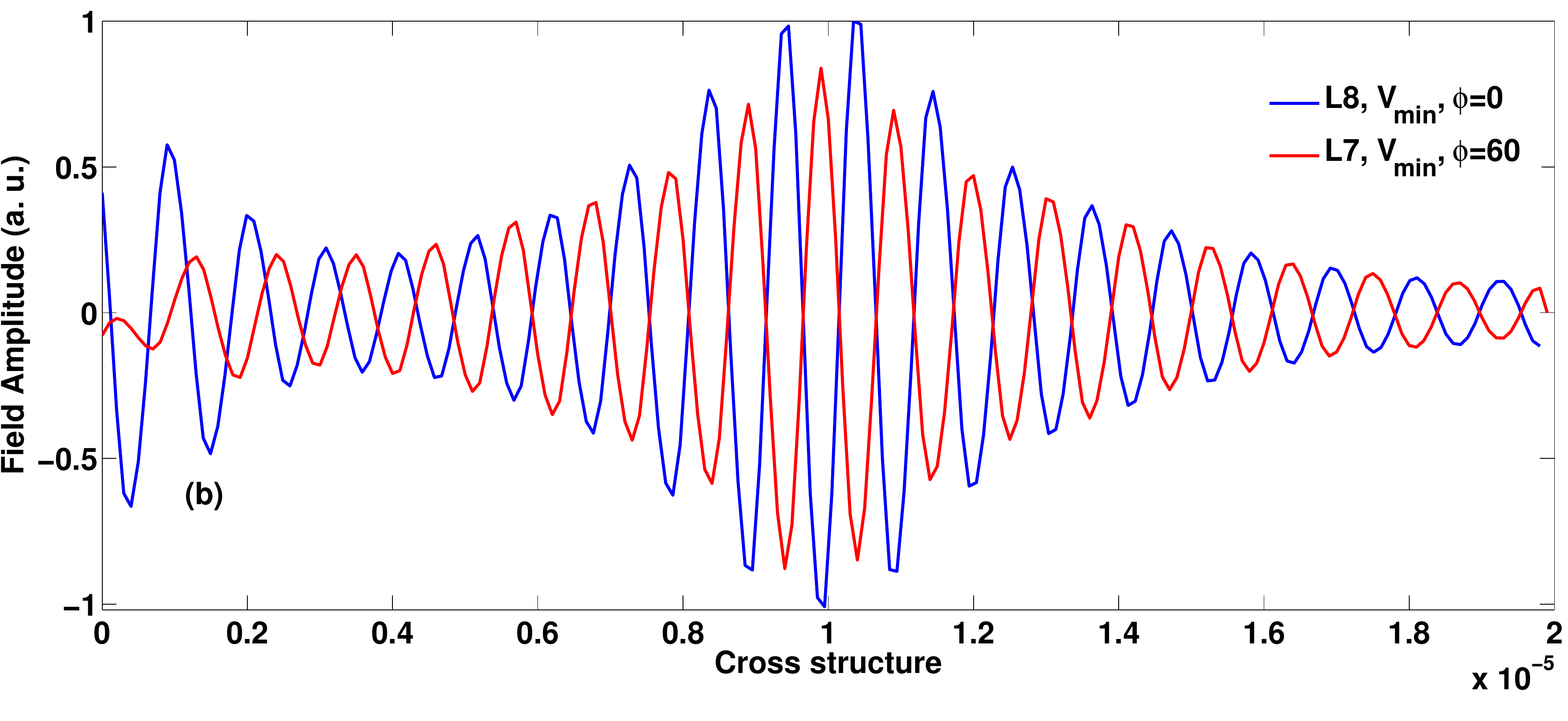}}
\centerline{
\includegraphics[height=3cm]{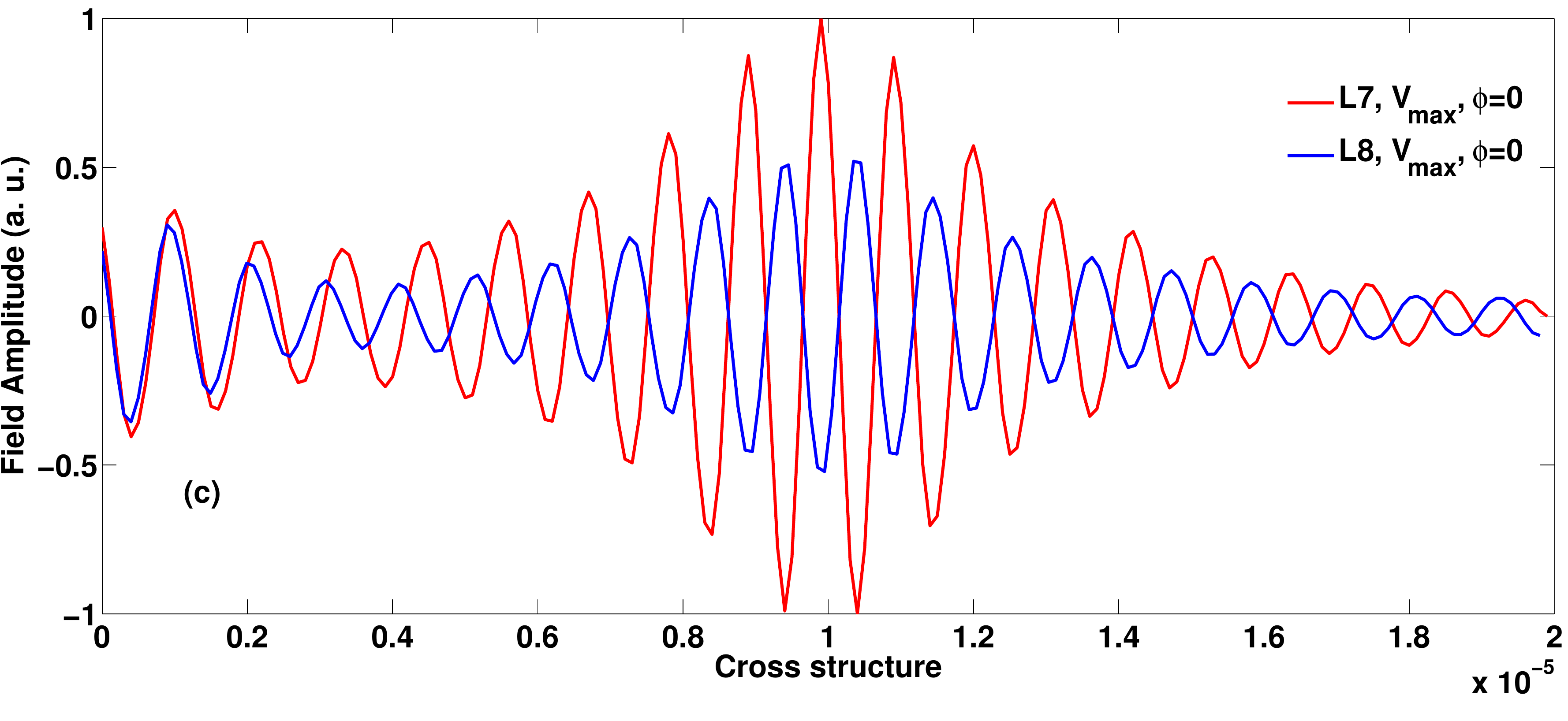}
\includegraphics[height=3cm]{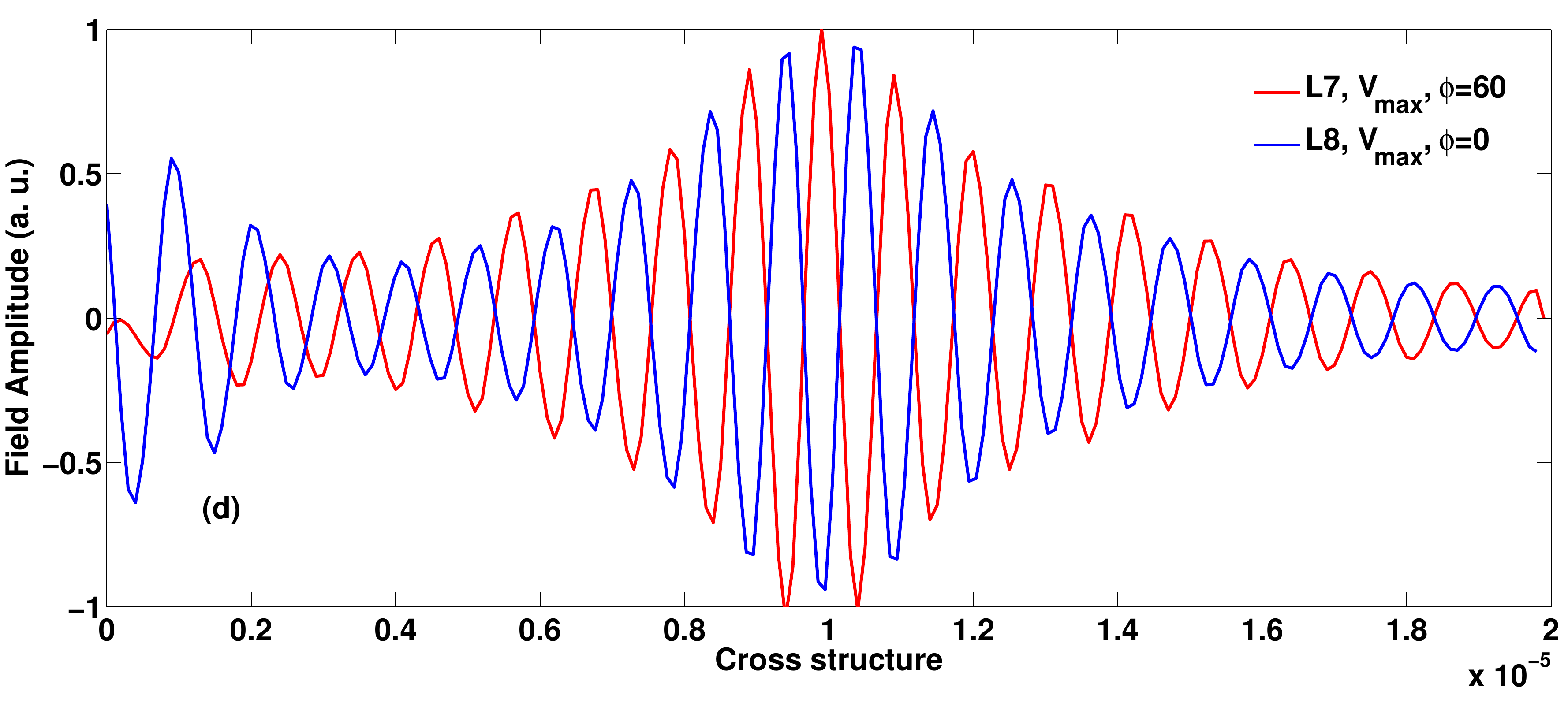}}
\caption{(Color online) Optical fields propagation through the
structure at the extreme wavelengthes, $1561.8 nm$ ($V_{min}$) and
$1538.8 nm$ ($V_{max}$), of the tunable rang for a and c) without
additional phase, i.e. $\phi_{iL7}= 0$ and $\phi_{iL8}= 0$; b and d)
with additional phase of $\phi_{iL7}=\pi/3$ while $\phi_{iL8}=0$.}
\label{fig3}
\end{figure*}

\begin{figure}[htb]
\begin{center}
\includegraphics[height=6cm]{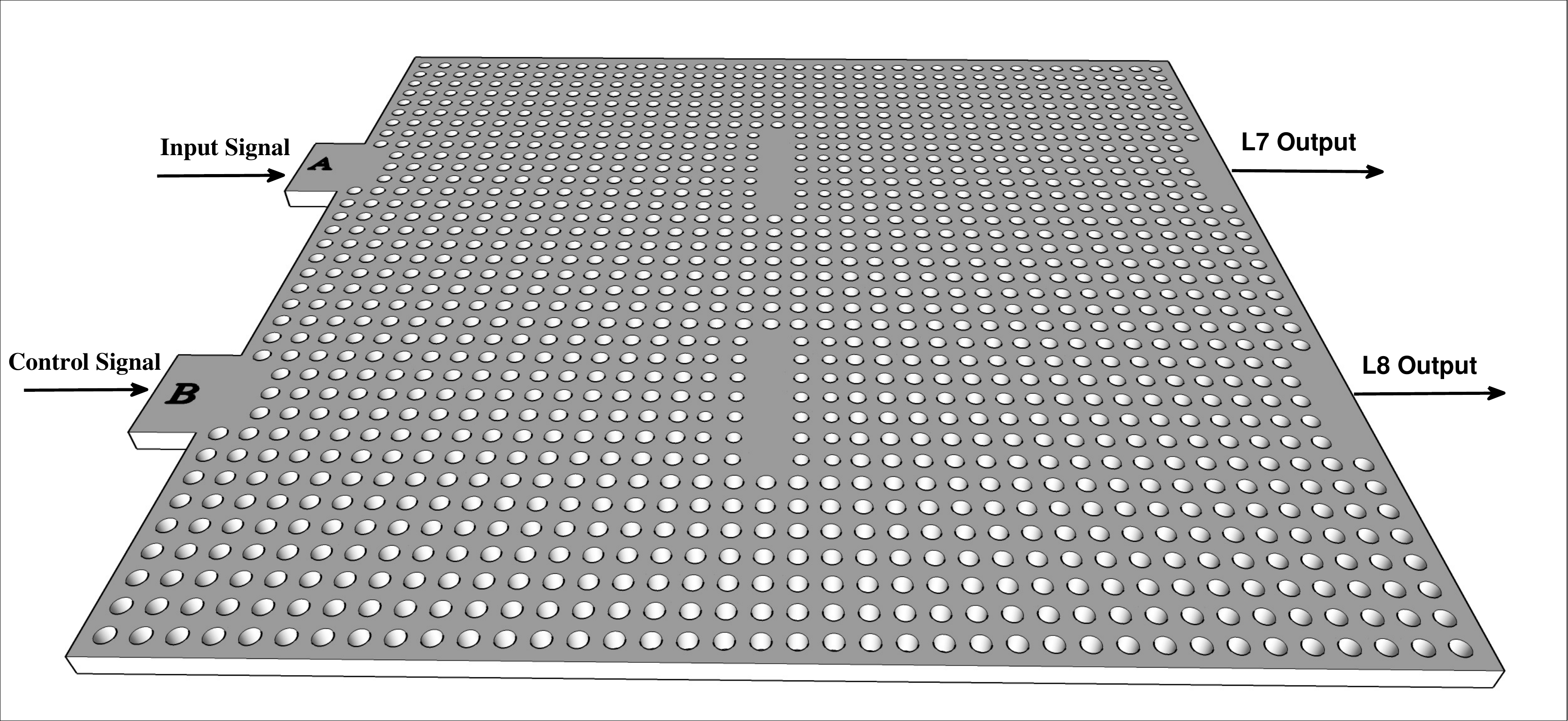}
\caption{(Color online) Schematic illustration of the CNOT (XOR)
gate. } \label{fig4}
\end{center}
\end{figure}

As indicated in table $1$, the $Q$ and $\lambda_r$ values are
different for each cavity type, which increase with increasing
cavity type. For the $L7$ cavity, the quality factor is equal to
$1472$ and for the $L8$ is $1816.86$. Also according to table $1$,
the resonance wavelength of the $L7$ and $L8$ cavities are $1561.1
nm$ and $1562.7 nm$, respectively.

By applying a single mode optical field to the $L7$ and $L8$
cavities, we can demonstrate that cavities resonate with $180$
degrees phase difference. Normalized transmission spectrum profile
for both of $L7$ and $L8$ cavity types are depicted in Figure $2$.
According to the Figure $2(a)$, it is observed that at wavelength
$1561.8 nm$, the transmitted electric field through $L7$ cavity has
a positive value while the transmitted electric field through $L8$
cavity has a negative value. We employed this transmission
characteristics of the cavities to propose ultra-fast all optical
CNOT (XOR) and XNOR gates. The resonant optical fields within the
$L7$ and $L8$ cavities experience a $\pi$ radian phase difference by
applying a Gaussian pulse with $1561.8 nm$ central wavelength. The
obtained phase difference cannot be retained through propagation in
$PhC$ due to the material impaction. Then, the phase difference
changes from $\pi$ radian to about $0.83pi$ ($150$ $deg.$) at the
output. To compensate this phase difference reduction, we applied
the input signal of the $L7$ cavity with a $\pi/3$ radian extra
phase. According to the observed Gaussian output pulses, we can see
that the proposed structure operate at very high speed ($fs$) as
illustrated in Figure $3$. In Figure $2(a)$, one can observe that
the resonance wavelengthes of cavities blue shift by increasing the
applied voltage. Where, a wavelength tunable range from $1561.8nm$
to $1538.8nm$ has been achieved by the applied voltage increasing
from $0$ to $3$ $V$. Figure $4$ shows the optical intensity of
signals with and without the extra phase through the structure.
Figures $4(a)$ and $(c)$ show evolution of the phase difference of
cavities from $180$ to $150 deg.$ through the structure in the cases
$V_{min}$ and $V_{max}$, respectively, for without the extra phase
case.

However, in the case of extra phase application
($\phi_{iL7}=\pi/3$), Figures $4(b)$ and $(d)$ illustrate the
evolution of the optical fields amplitudes for the cavities through
the structure in the cases $V_{min}$ and $V_{max}$, respectively, as
it is observed the phase difference evolution reach to $180 deg.$ in
the output of structure. With comparing Figures $4(a, c)$ and $4(b,
d)$ one can demonstrate that $180$ $deg.$ phase difference can be
obtained in output, which is useful for realization of the claimed
tunable optical logic gate.

\section{CNOT and XNOR logic gates}
\subsection{CNOT (XOR) gate}
To design a CNOT optical logic gate, we used two $L7$ and $L8$
cavities in $2D PhC$ with square lattice of air rods. We applied an
input signal to the $L7$ cavity with phase of $\phi_{iL7}=\pi/3 $ as
input for the CNOT gate and another input signal applied to the $L8$
cavity with input phase of $\phi_{iL8}= 0$ as the control signal for
the gate. Where, the field amplitude of control and input signals
are equal. When the control field is set in zero $"0"$ logic level,
with no signal applied to the $L8$ cavity, the input signal of the
$L7$ cavity propagates to the output without any considerable
reduction in its amplitude, as depicted in Figures $6(e, f)$ within
temporal range of $12fs$ to $24fs$. However, when the control field
is set in the high $"1"$ logic level, the reached pulse to the
output of the gate is inverted of the input signal as depicted in
Figures $6(e, f)$ from $0fs$ to $12fs$. Briefly, one can claim that
no variation in the signal pulse amplitude is produced without
applying any control signal. However, by applying a control signal,
a $\pi$ radian phase difference is produced between the input and
control signals, which results in a destructive interference at the
output and disappearing of the output signal referred as the low
$"0"$ logic level. As illustrated by Figure $5$, the input of $L7$
cavity is considered as the input port and the input of $L8$ cavity
as the control input of the CNOT gate. The central wavelength of the
both input pulses was the same at $1561.8 nm$. As explained in
above, Figures $6(e, f)$ exhibits the cavities responses with
Gaussian input pulses application for performing all logical cases
of the truth table for the CNOT logic gate. As an alternative view,
one can consider the performance of the illustrated structure as an
XOR gate, due to the wavelength and field amplitude sameness of the
applied input and control signals. Here, the evaluated least ON to
OFF logic-level contrast ratio for the XOR logic gate is calculated
as $25.45 dB$.

\subsection{XNOR gate}
In order to realize an XNOR gate, we should utilize three input
signals where each of them is applied for one of three cavities
created in the $2D PhC$. Here, there are two $L7$ cavities to apply
the input signals of the logic gate and one $L8$ cavity for applying
the control signal. For the XNOR gate, always, the control field is
set in high $"1"$ logic level. When the input signals are set in low
$"0"$ logic levels, without applying any input signal to any $L7$
cavities, the applied control signal to the $L8$ cavity propagates
to the output without any considerable variation in its amplitude,
as exhibited in Figures $7(g, h)$ within temporal range of $18fs$ to
$24fs$. However, when any of the input signals is set in the high
$"1"$ logic level, the received pulse to the output of the gate
becomes zero, as depicted in Figures $7(g, h)$ from $6fs$ to $18fs$.
As the last case, when both of the input signals are set in the high
$"1"$ logic level, interference of them with the control field cause
in emerging an output pulse in the high $"1"$ logic level as
exhibited in Figures $7(g, h)$ within the range of $0fs$ to $6fs$.
As the same as XOR, the evaluated least ON to OFF logic-level
contrast ratio for the XNOR logic gate is obtained as $22.61 dB$. In
this work, we have considered full width at half maximum ($FWHM$) of
the applied Gaussian signals equal with $1fs$. Where, we have
calculated the bit rate limit of $0.166$ $peta$ $bit$ $per$ $second$
$(pbps)$ for temporal performance of the simulated logic gates. Then
the proposed gates could demonstrate acceptable response for the
Gaussian input pulses with $FWHM$ value of $0.1fs$, temporally.
Thus, it is reasonable that data flow in the proposed logic gates
can be enhanced to ultrahigh rates of $0.166 pbps$, approximately.
In this section, we demonstrated the logic gates operation with
minimum value of the applied voltage at the central wavelength of
$1561.8 nm$. According to the Figure $2(a)$ we simulated operation
of the gates at both extreme wavelengths of the achieved $23 nm$
tunable range. Thus, the designed tunable logic gates based on the
electro-optic property of the substrate material show suitable
characteristics. Also, we would mention that one can design optical
tunable logic gates with employing only $L7$ or $L8$ cavities by
applying different voltages for each of the cavities.

\section{Conclusion}
The electro-optic property of the substrate material of the $2D PhC$
cavities is deployed in optical tunable CNOT (XOR) and XNOR gates
functions achievement. Where, we applied Gaussian pulse signals at
the central wavelength of $1561.8nm$ as input signals for $L7$ and
$L8$ cavities. Where, the resonance wavelength of the cavity can
reduce from $1561.8nm$ to $1538.8nm$ by increasing the applied
voltage to the each cavity. At any wavelength within the tunable
range, a $\pi$ radian phase difference is produced between the
cavities output signals. In summary, for retaining this phase
difference at the output of the $PhC$ structure, as the logic gate
output, we got need for an extra phase difference $(\pi/3)$ between
input and control signals of the logic gate. Also, by temporal
simulation of the designed logic gates, we demonstrated an
ultra-fast logic operation and show their high potential as an
appropriate candidate for applications such as tunable optical
integrated circuits and optical processors.

\markboth{Taylor \& Francis and I.T. Consultant}{Journal of Modern
Optics}

\begin{figure*}[htb]
\centerline{
\includegraphics[height=3cm]{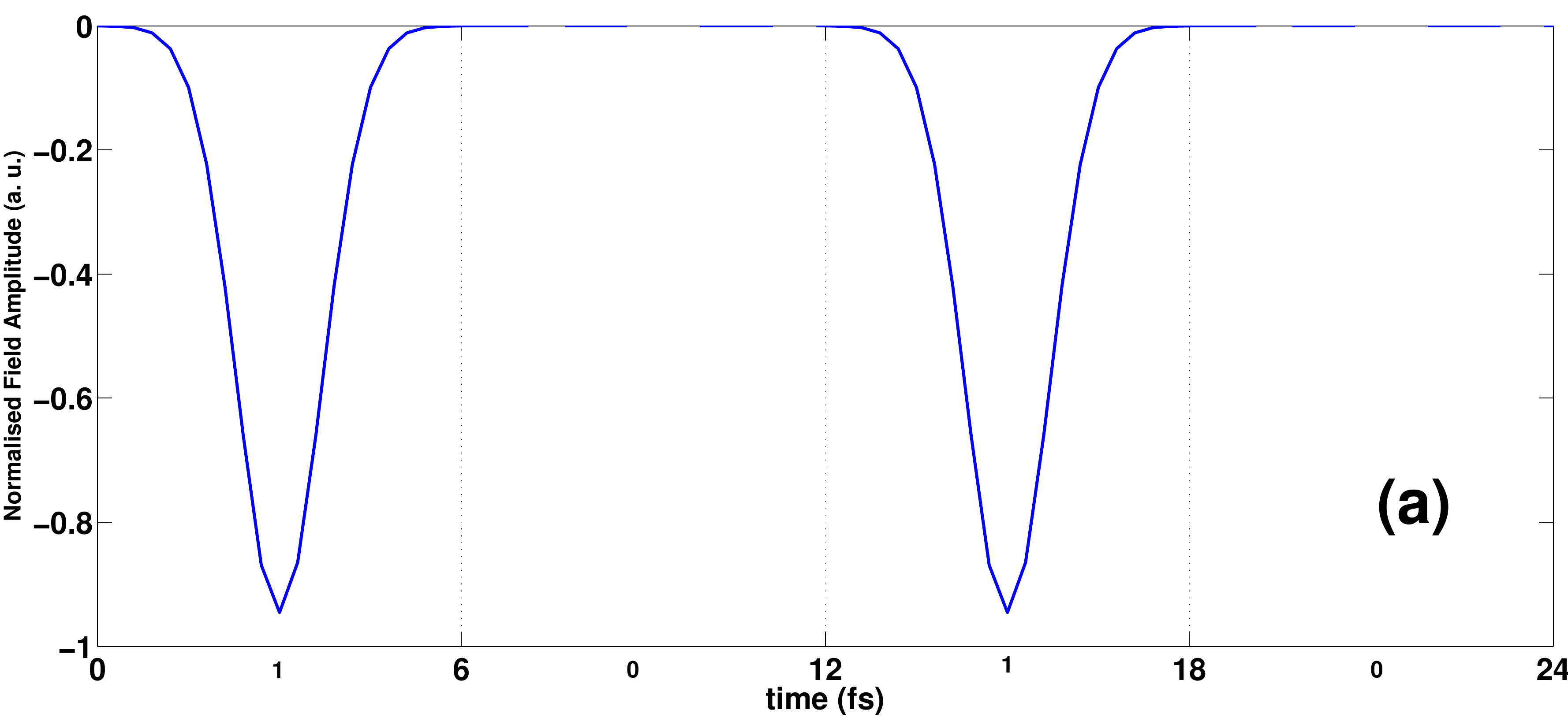}
\includegraphics[height=3cm]{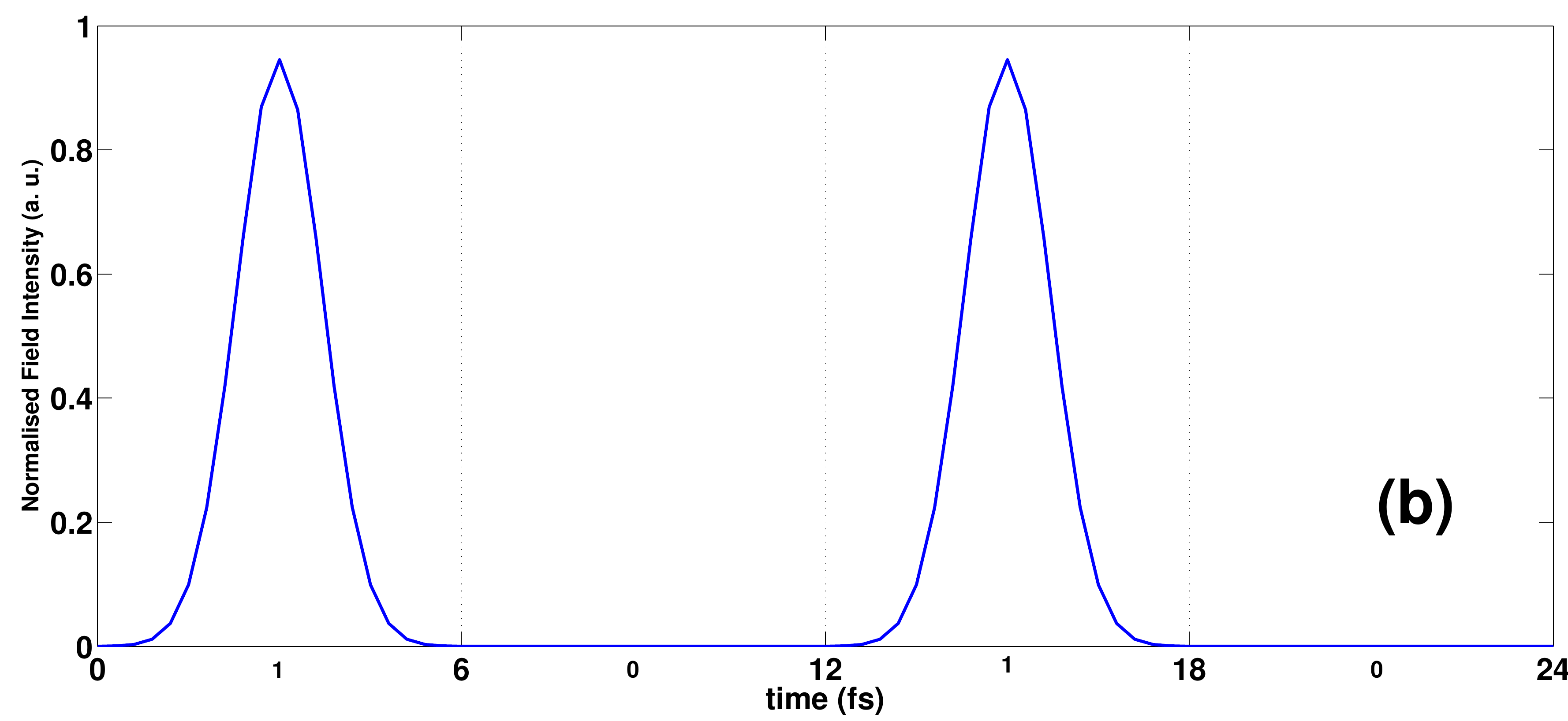}}
\centerline{
\includegraphics[height=3cm]{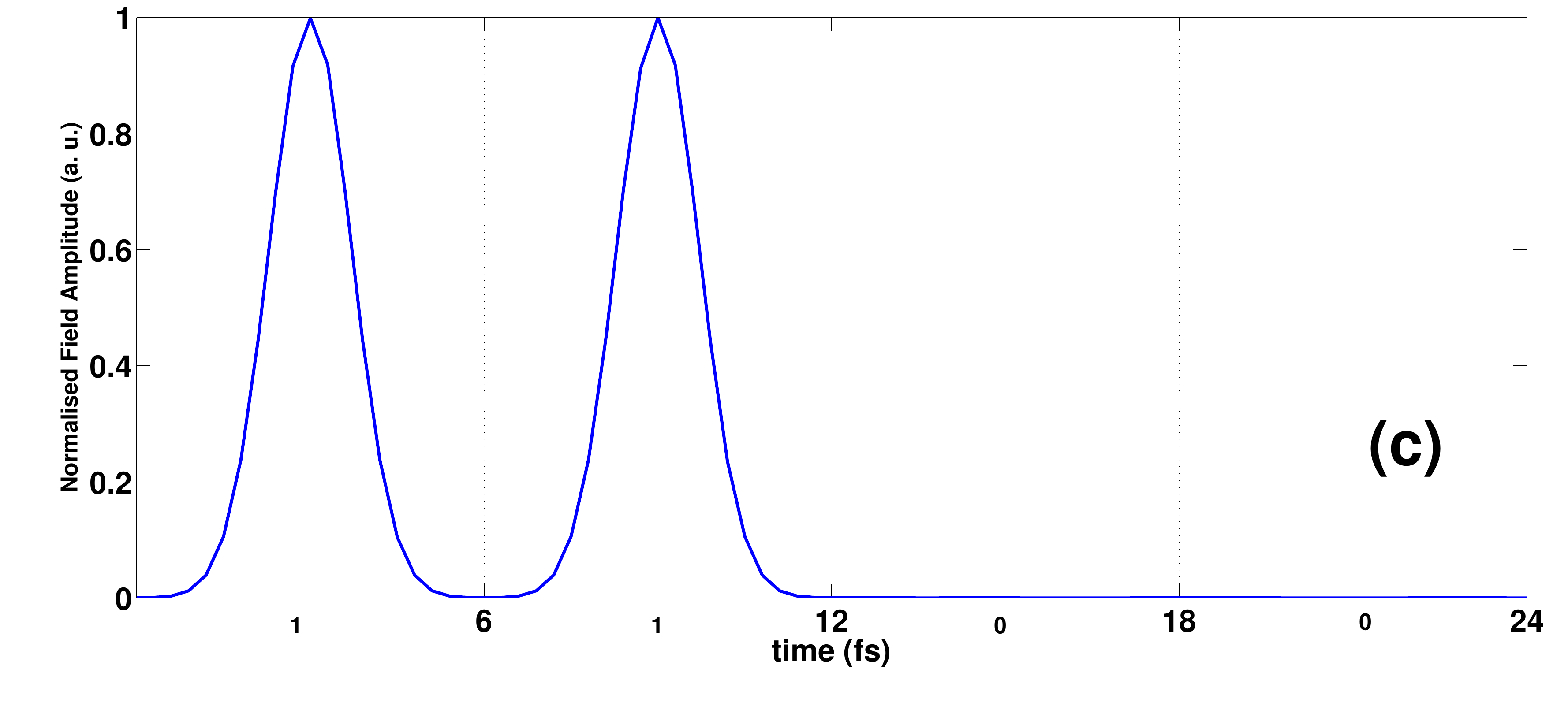}
\includegraphics[height=3cm]{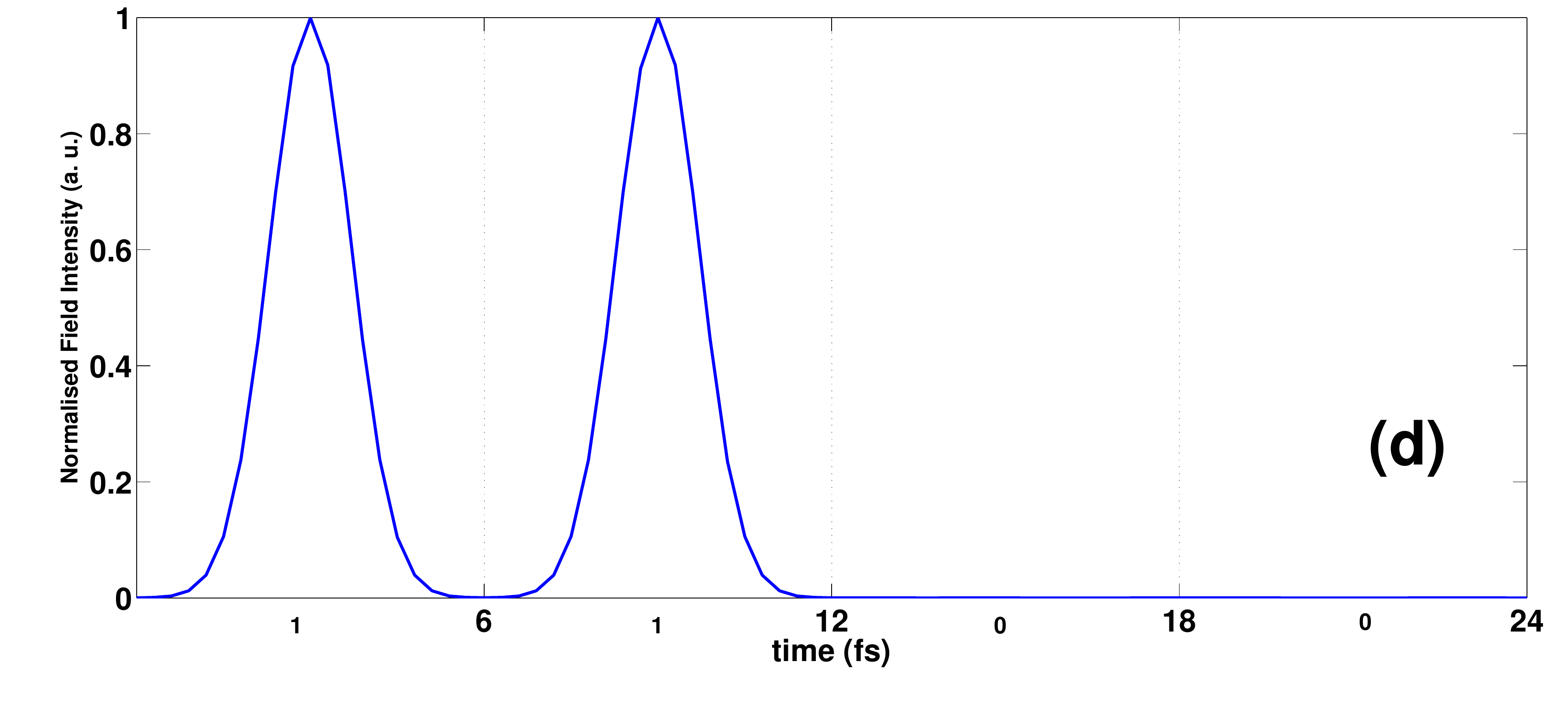}}
\centerline{
\includegraphics[height=3cm]{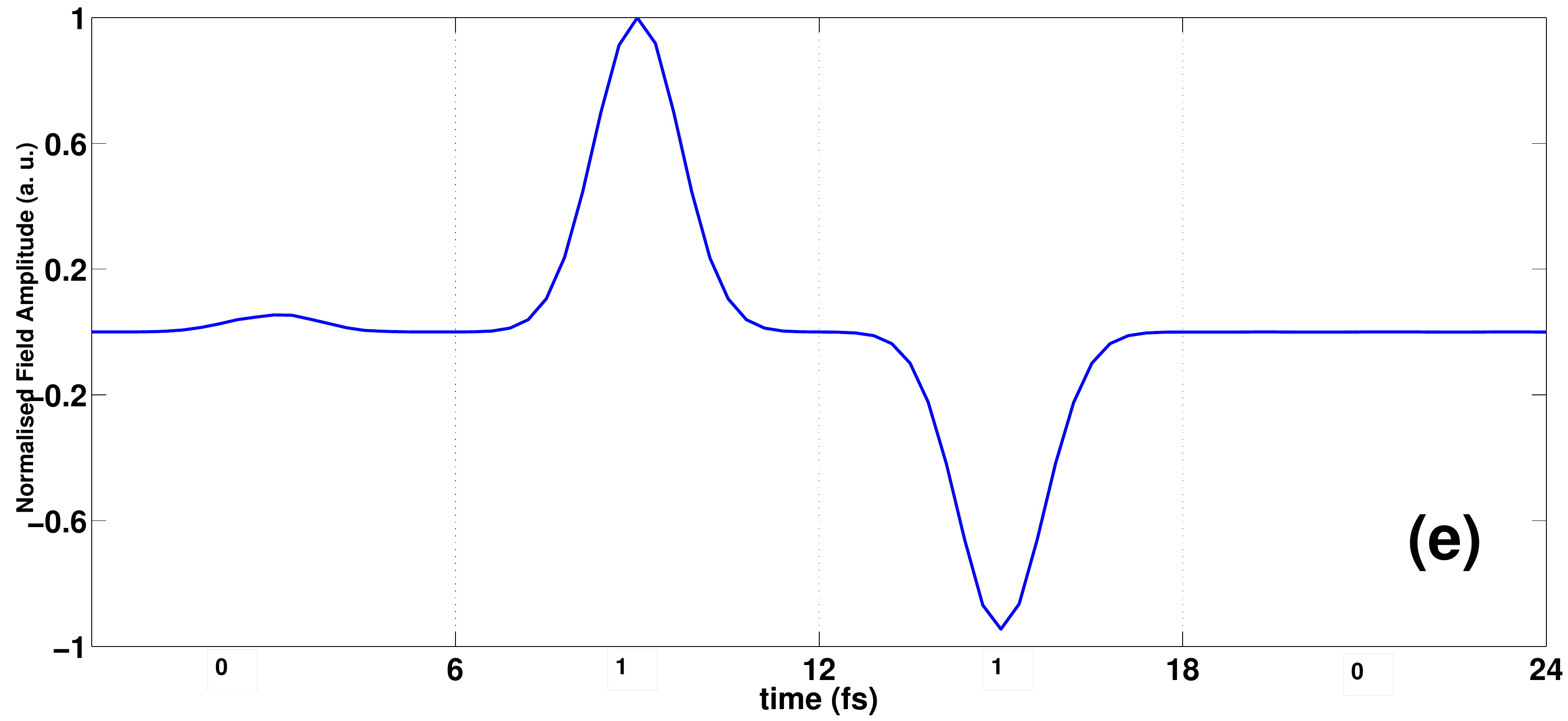}
\includegraphics[height=3cm]{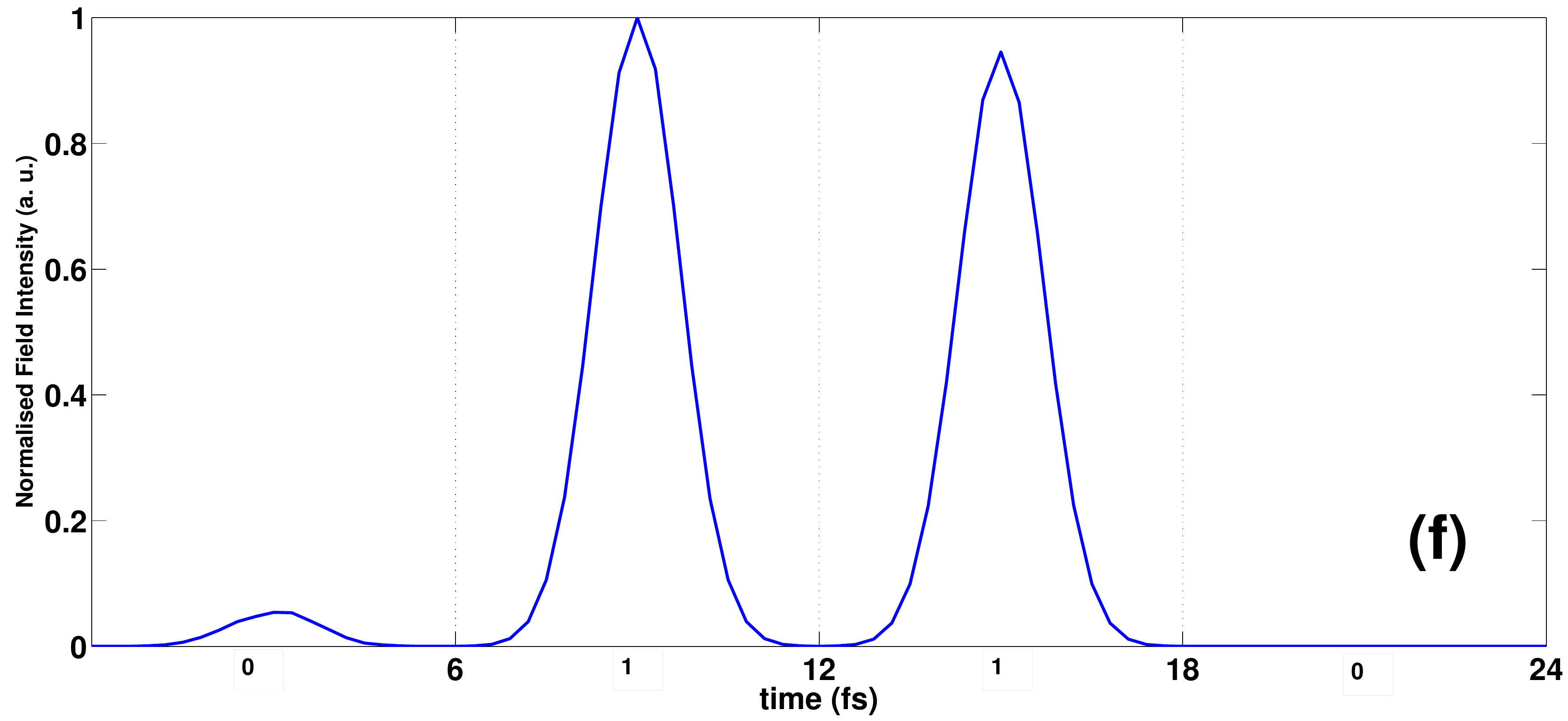}}
\caption{(Color online) Temporal view of the inputs and output
signals for the CNOT (XOR) gate; a and b) amplitude and intensity of
the input signal, c and d) amplitude and intensity of the control
signal, e and f) amplitude and intensity of the output signal.}
\label{fig3}
\end{figure*}

\begin{figure*}[htb]
\centerline{
\includegraphics[height=3cm]{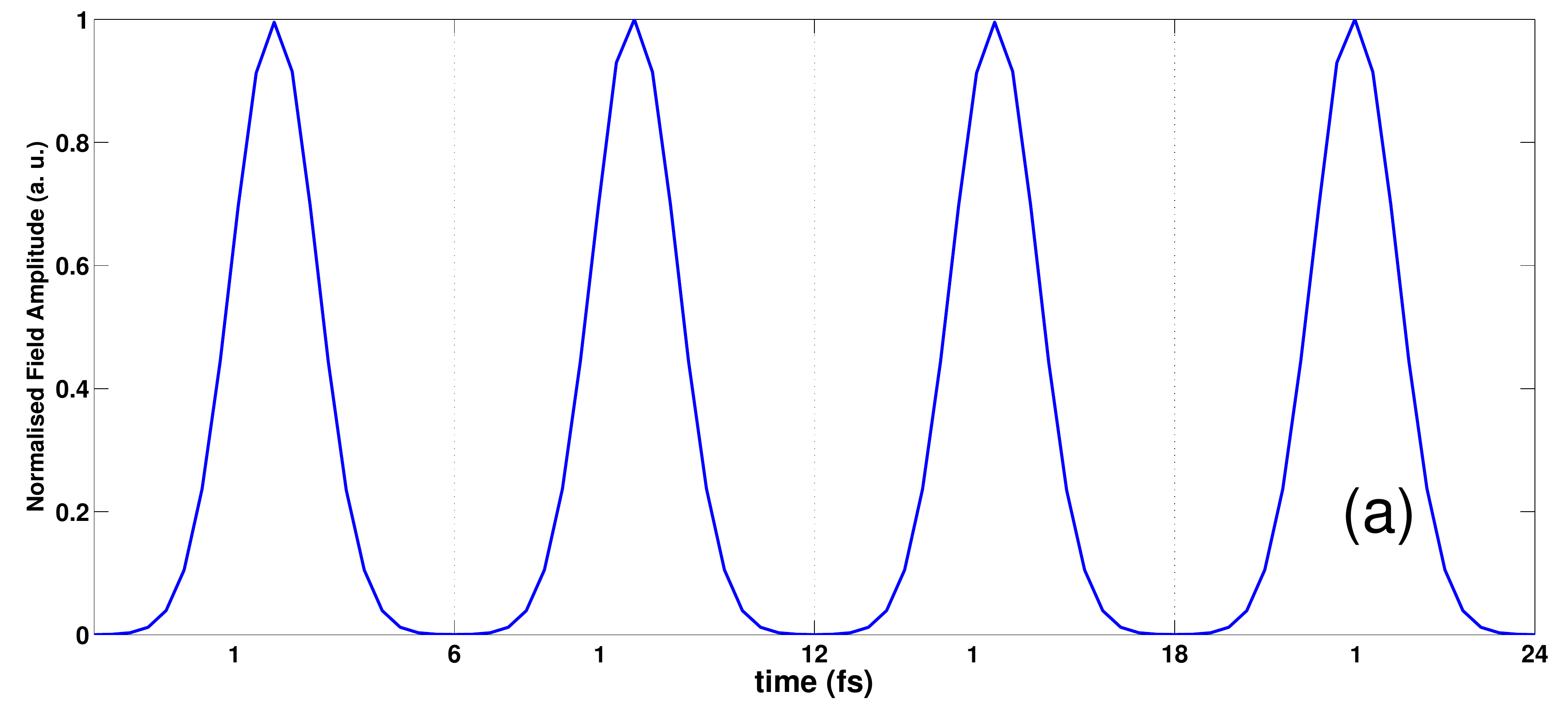}
\includegraphics[height=3cm]{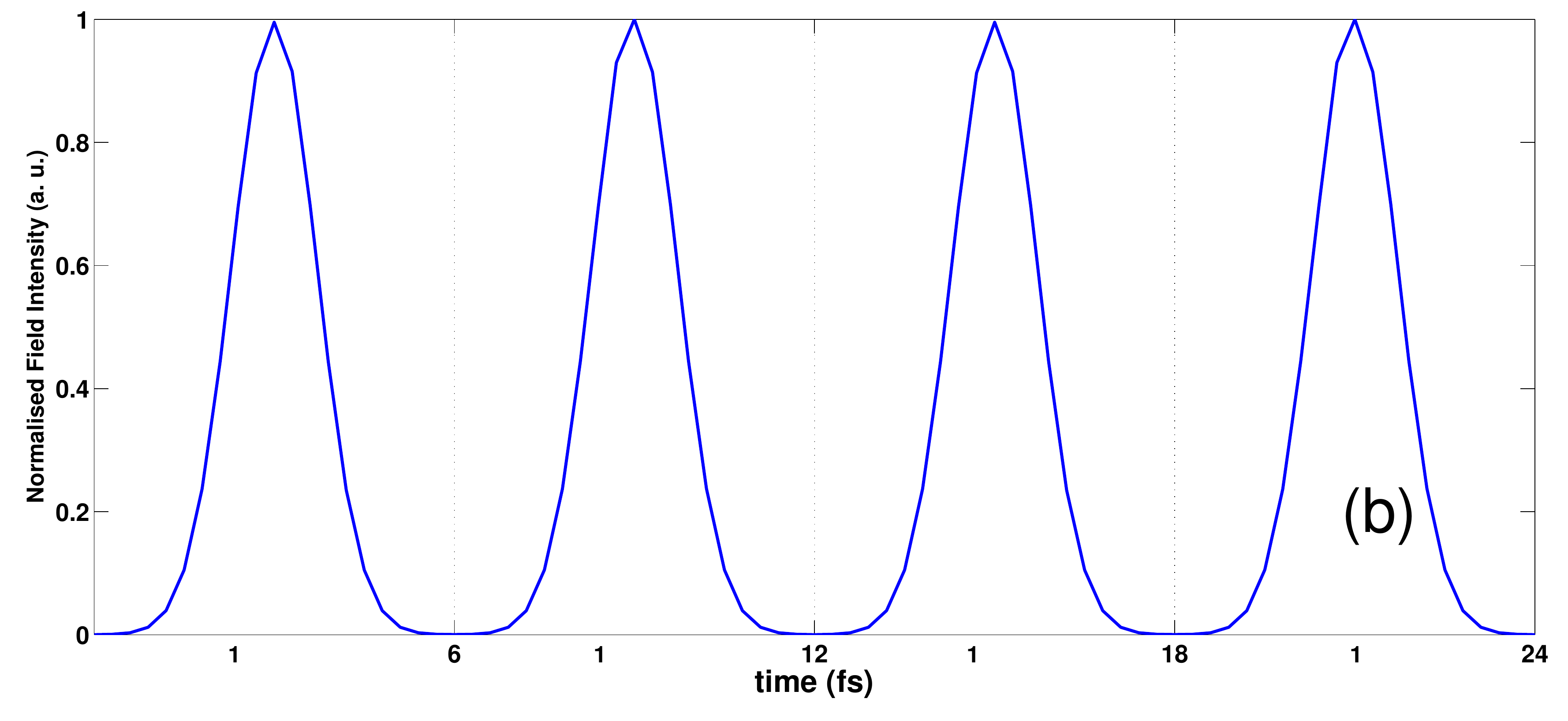}}
\centerline{
\includegraphics[height=3cm]{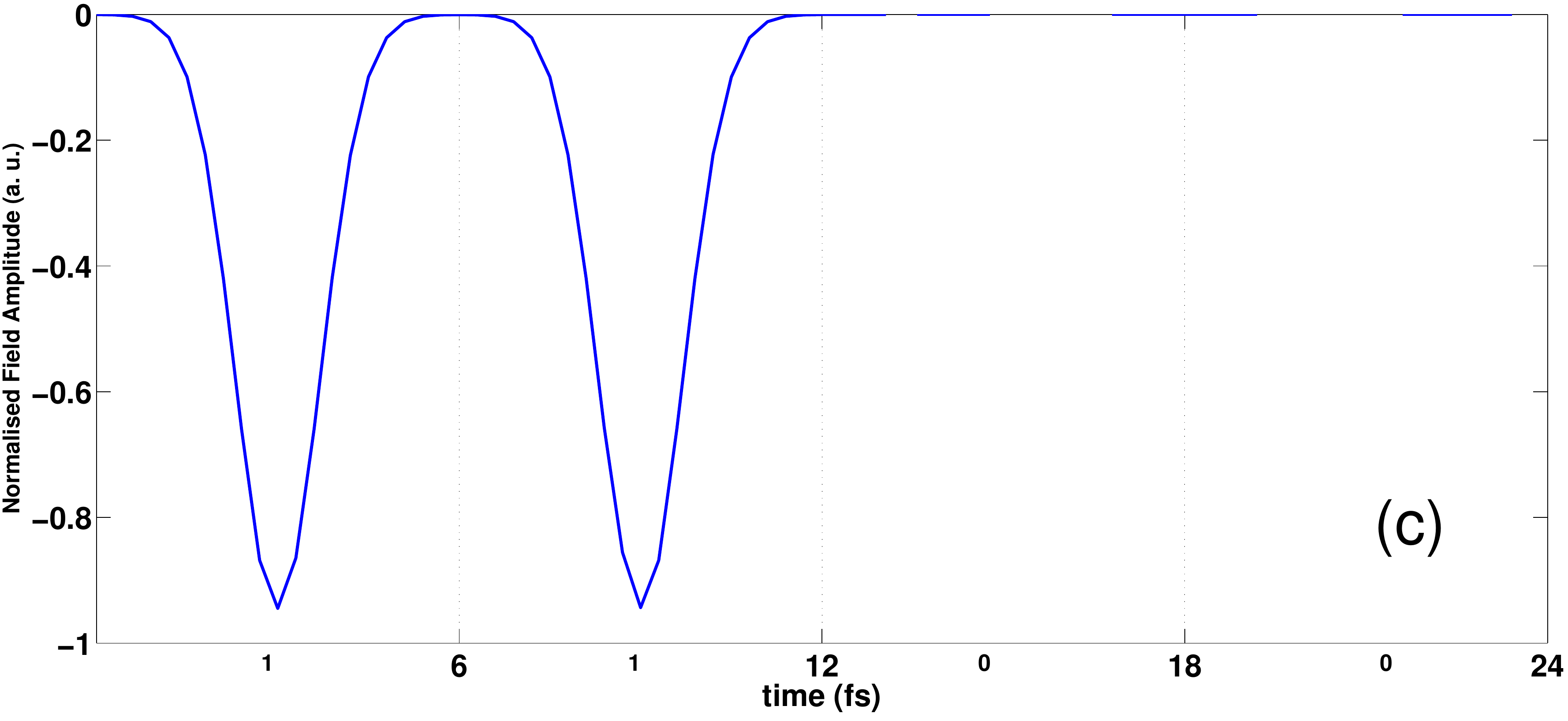}
\includegraphics[height=3cm]{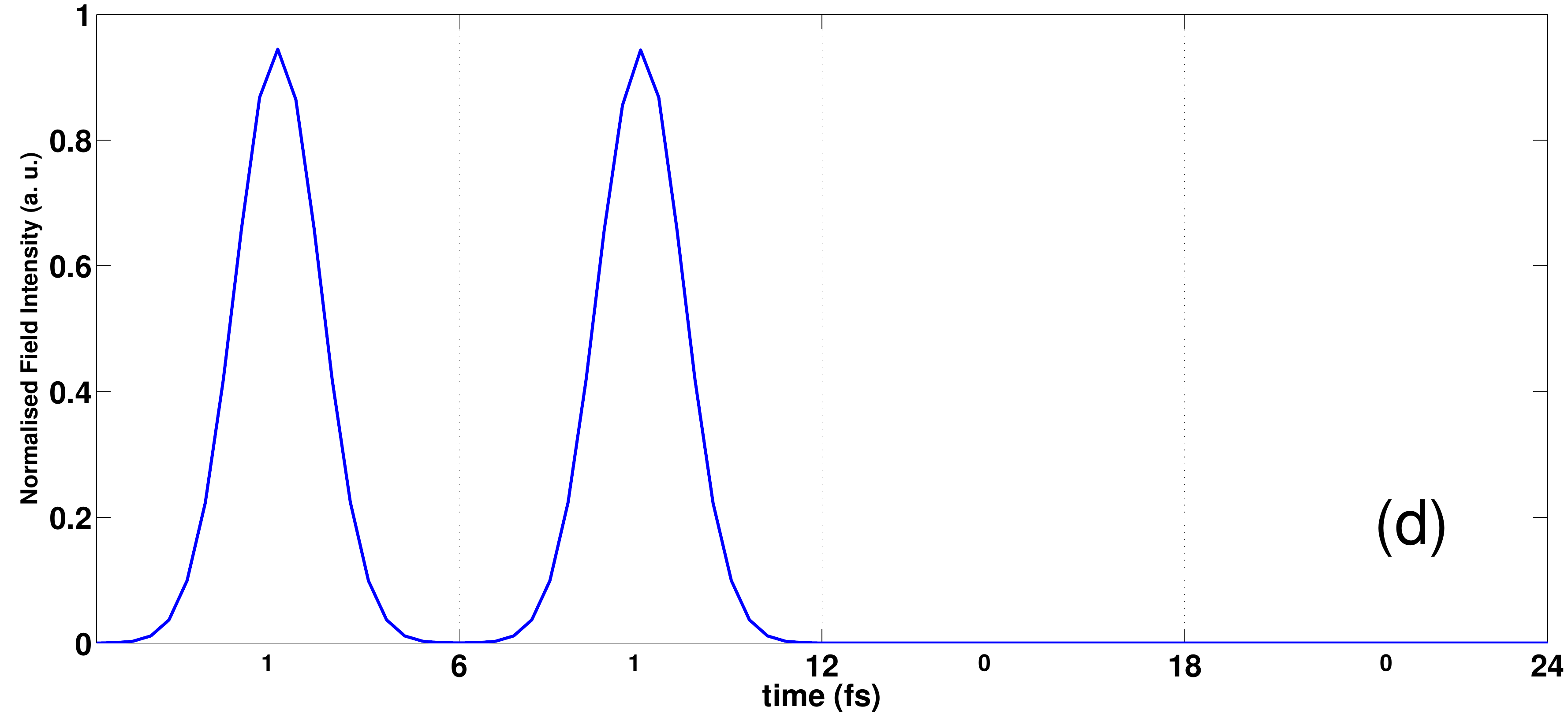}}
\centerline{
\includegraphics[height=3cm]{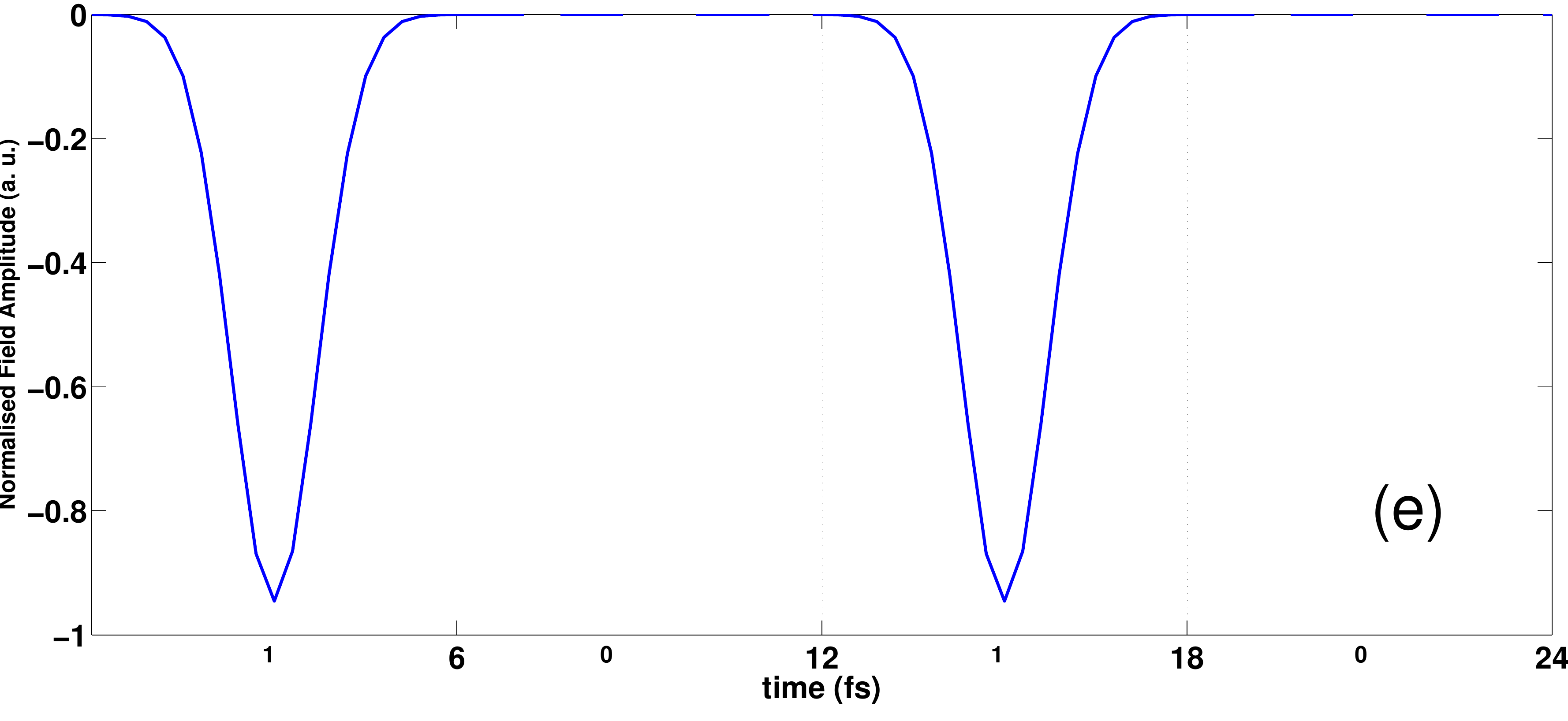}
\includegraphics[height=3cm]{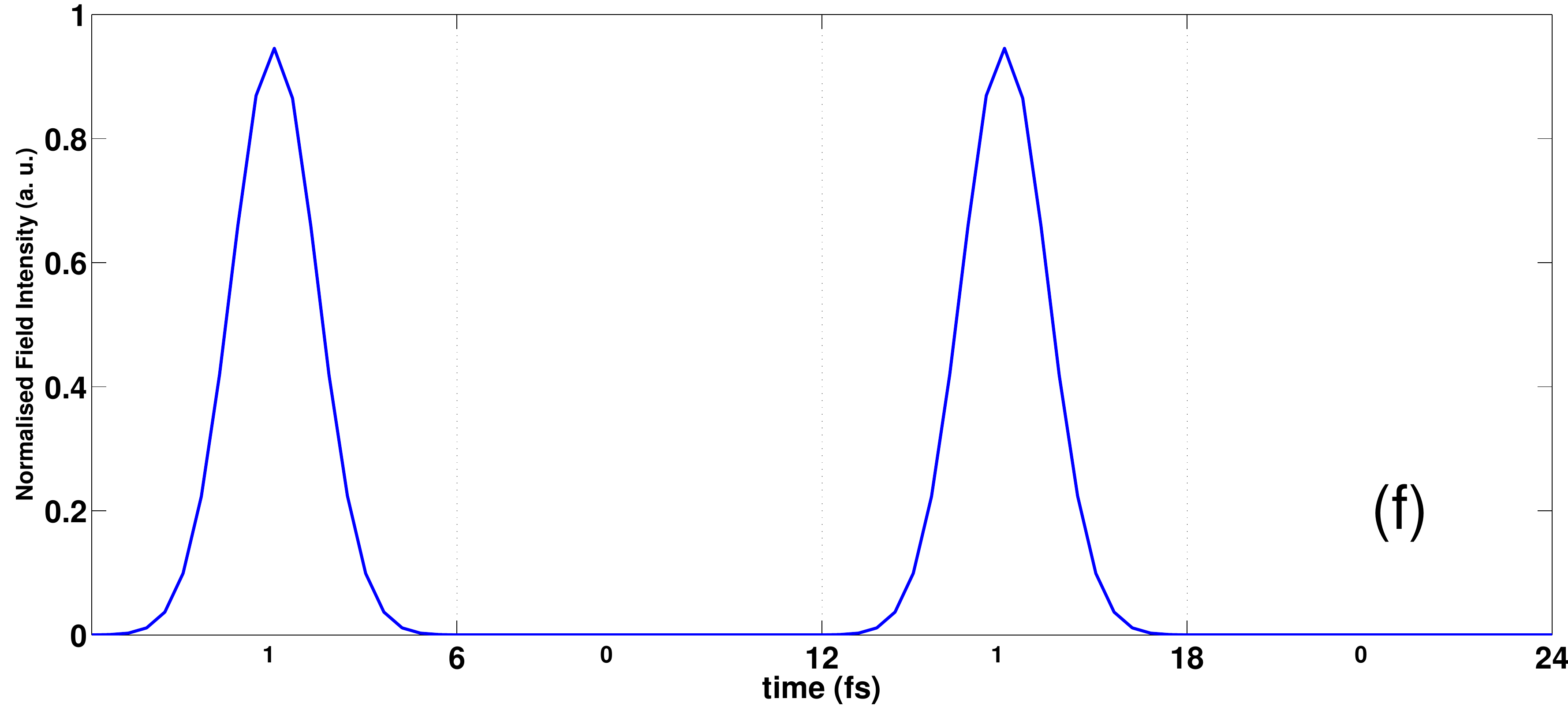}}
\centerline{
\includegraphics[height=3cm]{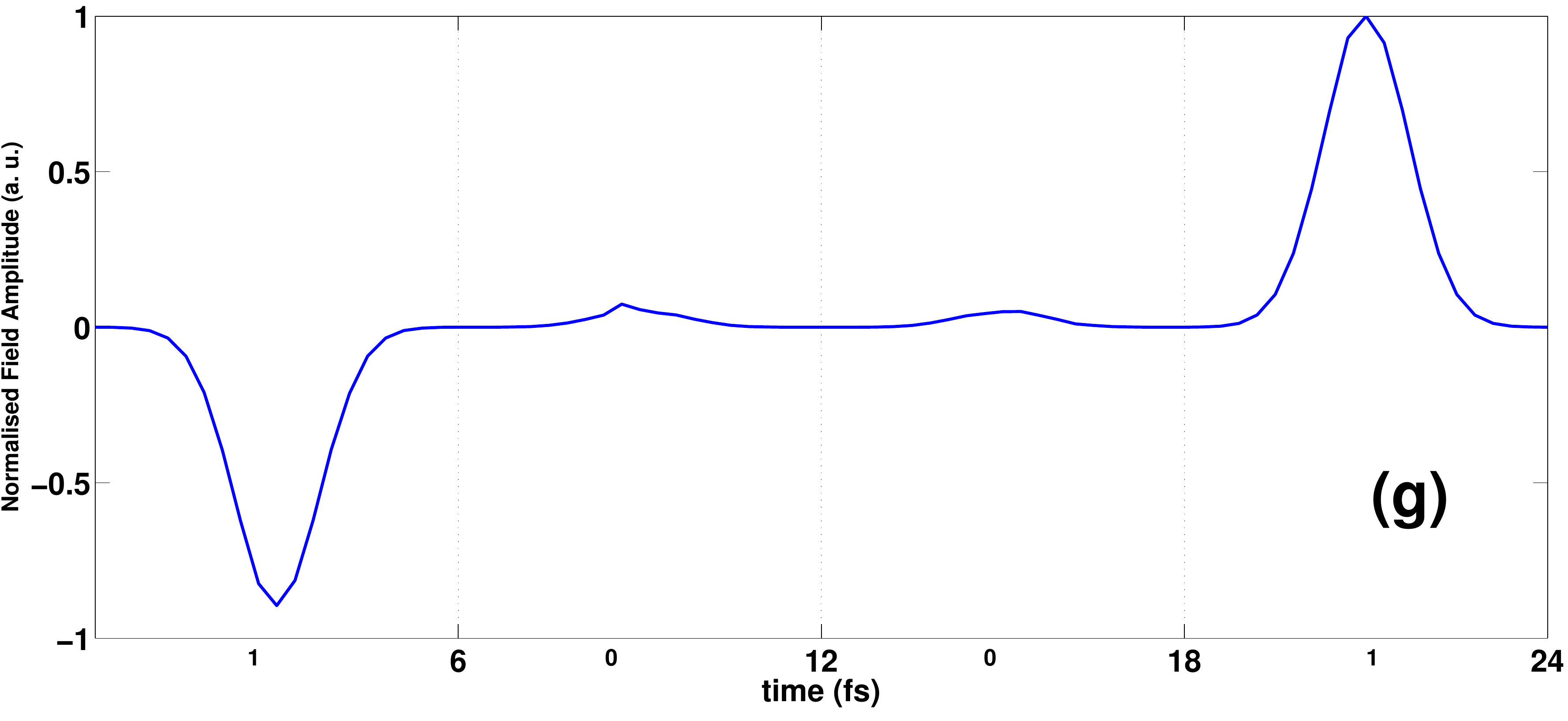}
\includegraphics[height=3cm]{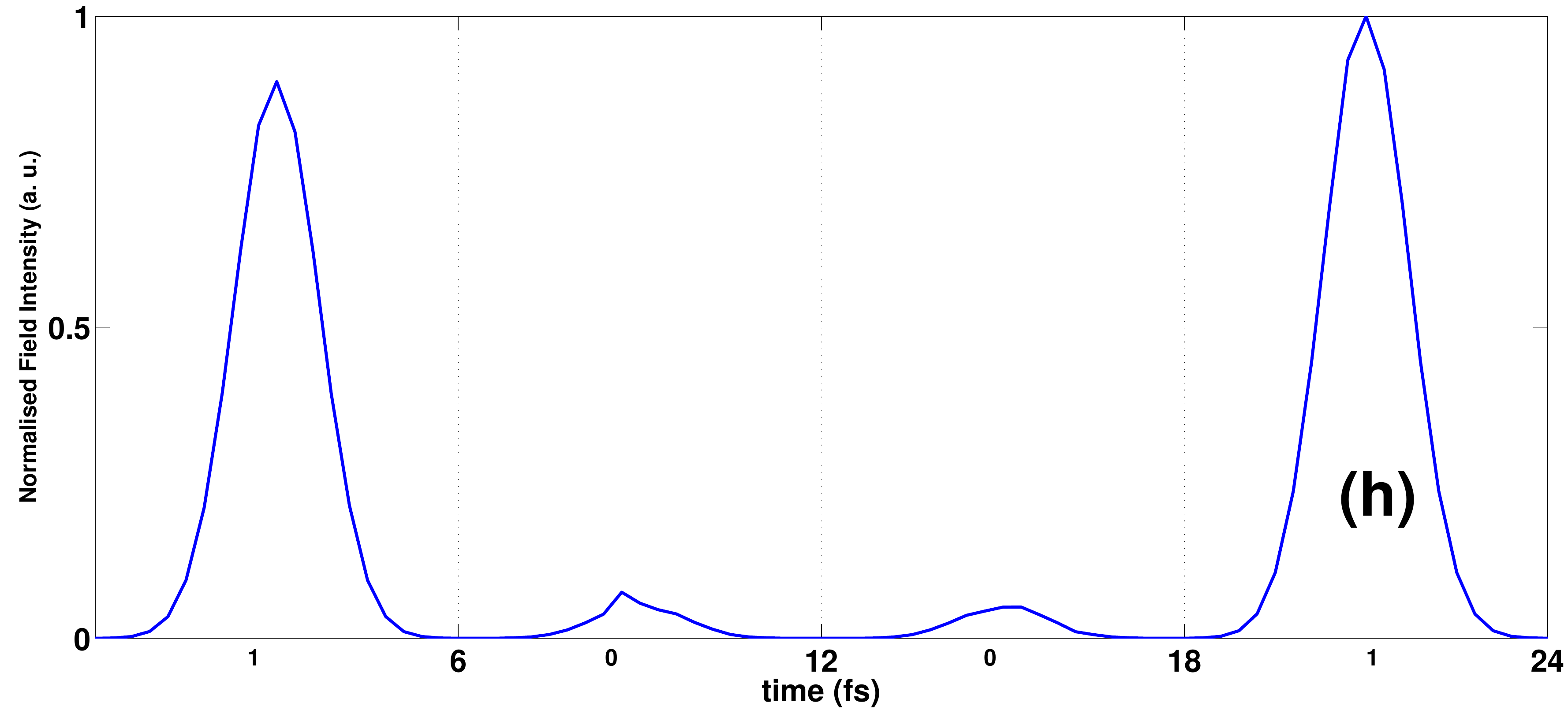}}
\caption{(Color online) Temporal view of the concerned signals in
the XNOR gate; a and b) amplitude and intensity of the control
Field, c and d) amplitude and intensity of the first input signal, e
and f) amplitude and intensity of the second input signal, g and h)
amplitude and intensity of the output signal.} \label{fig3}
\end{figure*}

\end{document}